\documentclass[draft]{article}

\newtheorem{prop}{Proposition}

\newcommand{\qed}{\rule{0.65em}{0.65em}}
\newenvironment{proof}[1][{}]{{\it Proof#1.\
}}{\hfill\qed\vspace{.5\baselineskip}} 
\newcounter{plist}
\renewcommand{\theplist}{\roman{plist}}
\newenvironment{proplist}{\begin{list}{(\theplist)}{\usecounter{plist}\setlength{\labelwidth}{5ex}}}{\end{list}}

\begin{document}

\author{A. Constandache\thanks{email: {\tt alexc@pas.rochester.edu}}\
\ and Ashok Das\thanks{email: {\tt das@pas.rochester.edu}}\\Department of
Physics and Astronomy\\University of Rochester\\Rochester, NY 14627-0171,
USA\\ \\ F. Toppan \thanks{email: {\tt
toppan@cbpf.br}}\\Centro Brasileiro de Pesquisas F\'{\i}sicas\\
Rua Xavier Sigaud 150 - Urca\\ 22290-180, Rio de Janeiro, RJ, Brasil} 
\title{Lucas polynomials and a standard Lax representation for the
polytropic gas dynamics} 
\maketitle
\begin{abstract}
A standard Lax representation for the polytropic gas dynamics is
derived by exploiting various properties of the Lucas and Fibonacci
polynomials. The two infinite sets of conserved charges are derived
from  this representation and shown to coincide with the ones derived
from  the known non-standard representation. The same Lax function is
shown to also give the standard Lax description for the elastic medium
equations. In addition, some results on possible dispersive extensions
of  such models are presented.
\end{abstract}
\pagebreak

\section{Introduction:}

Systems of hydrodynamic type \cite{1}-\cite{5} have been studied
extensively  over the
last several years. Such systems manifest, among others, in string
theories, membrane theories and topological field theories \cite{6}. The
polytropic gas equations \cite{7} belong to this general class of
systems  and are described by
\begin{equation}
u_{t} + uu_{x} + v^{\gamma -2} v_{x} = 0,\qquad v_{t} + (uv)_{x} = 0
\end{equation}
where $u,v$ denote the two dynamical variables. For different values of
the exponent, $\gamma$, these equations describe different physical
systems, which are dispersionless and integrable \cite{8}.

Being dispersionless integrable systems, hydrodynamic systems can be
described by a Lax equation on the classical phase space \cite{8}. The Lax
description that has been obtained so far for the polytropic gas
equations \cite{9}-\cite{10}, however, is what is called
a non-standard representation, which is not very useful in
generalizing this system to other cases, such as the supersymmetric
polytropic gas equations \cite{11}. Nevertheless, the
non-standard Lax description has been quite useful
\cite{9}-\cite{10}. It has led
naturally  to the two infinite sets of conserved charges of the system
(by the standard construction, although the two sets are obtained by
expanding the residues at different points). The Lax description also
immediately leads to the involution of the charges and clarifies why
both the polytropic gas equations and the elastic medium equations
share the same set of conserved charges (basically because they are
both described by the same Lax function). One can also construct the
Hamiltonian structures from this description, although a more
convenient construction is through the Moyal-Lax representation of the
system \cite{12}. 

Many integrable systems can have both a standard as well as a
non-standard Lax representation and, as we have already mentioned, a
standard representation is much more useful. Nonetheless, such a
description is lacking so far and, in this paper, we construct such a
representation for the polytropic gas equations. Interestingly, such a
description involves the use of Fibonacci and Lucas polynomials and
their properties \cite{13}. Although a standard Lax representation has 
unique residues, we show that such a Lax
description, nevertheless, leads to the two infinite sets of conserved
charges coming from two distinct families of fractional powers of the
Lax  function. The same
Lax function also provides a standard Lax representation for the
elastic medium equations \cite{7,9,10}.

The paper is organized as follows. In section {\bf 2}, we review, very
briefly, the definitions and some essential properties of the
Fibonacci and Lucas polynomials \cite{13}. In section {\bf 3}, we
derive  some
identities satisfied by two auxiliary functions dependent on these
polynomials. Using these, we construct, in section {\bf 4}, the
standard Lax  representation
for the polytropic gas equations. In section {\bf 5}, we construct the
two  infinite sets of
conserved charges for this system, which coincide with the
earlier known results \cite{9}. The involution of these charges is automatic,
since they come from a Lax description. In section {\bf 6}, we attempt
to construct some dispersive equations, whose dispersionless limit may
lead to these equations. This question is extremely difficult and we
present only some partial results on this issue. We end with a brief
conclusion in section {\bf 7}. In the appendix, we show how the same
Lax function that leads to a standard Lax description for the
polytropic gas equations also describes the elastic medium
equations. This also shows that the two systems share the two infinite
sets of conserved charges.

\section{Definition and properties of the Lucas and Fibonacci polynomials:}
The Lucas polynomials are defined recursively as follows:
\begin{equation}\label{eq-1}
l_{n+1}(x)=x\,l_{n}(x)+l_{n-1}(x)
\end{equation}
with $l_{0}(x)=2$ and $l_{1}(x)=x$. Their explicit form for $n\geq 1$ is:
\begin{equation}\label{eq-2}
l_{n}(x)=\sum_{k=0}^{\lfloor\frac{n}{2}\rfloor}\frac{n}{n-k}
\left(\begin{array}{c}n-k\\k\end{array}\right)x^{n-2k} 
\end{equation} 
where $\lfloor x\rfloor$ is the largest integer smaller than or equal
to $x$  and
$\left(\begin{array}{c}n\\m\end{array}\right)$ is the binomial
coefficient. 

The Fibonacci polynomials are defined by the same recursion relation:
\begin{equation}\label{eq-3}
f_{n+1}(x)=x\,f_{n}(x)+f_{n-1}(x)
\end{equation}
but with $f_{0}(x)=0$ and $f_{1}(x)=1$. Their explicit form for $n\geq 1$ is:
\begin{equation}\label{eq-4}
f_{n}(x)=\sum_{k=0}^{\lfloor\frac{n-1}{2}\rfloor}
\left(\begin{array}{c}n-k-1\\k\end{array}\right)x^{n-2k-1}
\end{equation}

A good presentation of these two families of polynomials and the many
relations involving them can be found in \cite{13}. However, for
convenience, we list below some of the relations which we are going to use
throughout the paper: 
\begin{equation}\label{eq-5}
l_{n}(x)=f_{n+1}(x)+f_{n-1}(x)
\end{equation}
\begin{equation}\label{eqn-6}
(x^{2}+4)\,f_{n}(x)=l_{n+1}(x)+l_{n-1}(x)
\end{equation}
\begin{equation}\label{eqn-7}
l_{n}^{2}(x)-(x^{2}+4)\,f_{n}^{2}(x)=4(-1)^{n}
\end{equation}
\begin{equation}\label{eq-8}
l_{n}'(x)=n\,f_{n}(x)
\end{equation}

\section{Various useful identities:}
In addition to the Lucas and Fibonacci polynomials, the following two
families of polynomials are also useful in deriving some of our
results: 
\begin{equation}\label{eq-9}
g_{n}(x)=\sum_{k=0}^{\lfloor\frac{n}{2}\rfloor}\frac{n}{n-k}
\left(\begin{array}{c}n-k\\k\end{array}\right)x^{k}
\end{equation}
\begin{equation}\label{eq-10}
h_{n}(x)=\sum_{k=0}^{\lfloor\frac{n}{2}\rfloor}
\left(\begin{array}{c}n-k\\k\end{array}\right)x^{k}
\end{equation}
If $x\not = 0$, one can easily see from (\ref{eq-2}) and (\ref{eq-4}) that:
\begin{equation}\label{eq-11}
l_{n}(x)=x^{n}g_{n}\left(\frac{1}{x^2}\right)
\end{equation}
for $n\geq 0$ and:
\begin{equation}\label{eq-12}
f_n(x)=x^{n-1}h_{n-1}\left(\frac{1}{x^2}\right)
\end{equation}
for $n\geq 1$. Using this connection with the Lucas and Fibonacci
polynomials, we can prove that  
\begin{prop}\label{prop-1}
For any $n\geq 2$, the polynomials $g_{n}$ and $h_{n}$ satisfy the
following relations: 
\begin{proplist}
\renewcommand{\theenumi}{\roman{enumi}}
\item\label{prop-1-1} $h_{n}(z)=h_{n-1}(z)+z\,h_{n-2}(z)$
\item\label{prop-1-2} $g_{n}(z)=g_{n-1}(z)+z\,g_{n-2}(z)$
\item\label{prop-1-3}
$g_{n}(z)=h_{n}(z)+z\,h_{n-2}(z)=h_{n-1}(z)+2\,z\,h_{n-2}(z)$ 
\item\label{prop-1-4} $g_{n}'(z)=n\,h_{n-2}(z)$
\item\label{prop-1-5}
$g_{n+1}(z)-\frac{2}{n+1}\,z\,g_{n+1}'(z)=g_{n}(z)-\frac{1}{n}\,z\,g_{n}'(z)$ 
\item\label{prop-1-6}
$\frac{1}{n+1}\,g_{n}(z)\,g_{n+1}'(z)-\frac{1}{n}\,g_{n}'(z)\,g_{n+1}(z)=
(-z)^{n-1}$ 
\end{proplist}
\end{prop} 
\begin{proof}
To prove (\ref{prop-1-1}), we use (\ref{eq-12}) to rewrite (\ref{eq-3}) as
\begin{displaymath}
h_{n}\left(\frac{1}{x^2}\right)=h_{n-1}\left(\frac{1}{x^2}\right)+
\frac{1}{x^2}\,h_{n-2}\left(\frac{1}{x^2}\right) 
\end{displaymath}
for $x\not = 0$. If we now let $z=x^{-2}$ we find that
(\ref{prop-1-1}) holds for $z\not = 0$. Since both sides of
(\ref{prop-1-1}) are polynomial, it follows by continuity that it must
also hold at $z=0$. In a similar way (\ref{prop-1-2}) follows from
(\ref{eq-1}), (\ref{prop-1-3}) follows from (\ref{eq-5}) and
(\ref{prop-1-4}) follows from (\ref{eq-8}). Let us note from
(\ref{prop-1-3}) and (\ref{prop-1-4}) that 
\begin{displaymath}
g_{n+1}(z)-\frac{2}{n+1}\,z\,g_{n+1}'(z)=h_{n}(z)+2\,\,z\,
h_{n-1}(z)-2\,\,z\,h_{n-1}(z)=h_{n}(z) 
\end{displaymath}
as well as
\begin{displaymath}
g_{n}(z)-\frac{1}{n}\,z\,g_{n}'(z)=h_{n}(z)+z\,h_{n-2}(z)-z\,h_{n-2}(z)=
h_{n}(z)
\end{displaymath}
so that (\ref{prop-1-5}) holds.
Finally, for (\ref{prop-1-6}) we proceed by induction. For $n=2$ the
relation obviously holds. Now suppose that it holds for some $n$ and
notice that: 
\begin{eqnarray*}
& &\frac{1}{n+2}\,g_{n+1}(z)\,g_{n+2}'(z)-\frac{1}{n+1}\,
g_{n+1}'(z)\,g_{n+2}(z)\\
& &=g_{n+1}(z)\,h_{n}(z)-h_{n-1}(z)\,g_{n+2}(z)\\
&
&=g_{n+1}(z)\,(h_{n-1}(z)+z\,h_{n-2}(z))-h_{n-1}(z)\,(g_{n+1}(z)+z\,g_{n}(z))\\
& &=(-z)\,(h_{n-1}(z)\,g_{n}(z)-g_{n+1}(z)\,h_{n-2}(z))\\
&
&=(-z)\left(\frac{1}{n+1}\,g_{n+1}'(z)\,g_{n}(z)-\frac{1}{n}\,g_{n+1}(z)\,
g_{n}'(z)\right)\\
& &=(-z)^{n}
\end{eqnarray*}
The relation therefore holds for $n+1$, which completes the proof.  
\end{proof}

Let $\phi(z)$ be a function which is analytic in a disc
$D=\{z:|z|<r\}$ and let 
$$
\phi(z) = \sum_{n=0}^{\infty}\,\gamma_{n}\,z^{n}
$$ 
be its
Taylor expansion around $z=0$.  We will denote by $[\phi]_{m}(z)$ the
polynomial consisting of terms in the Taylor expansion 
of $\phi$ around $z=0$ up to $z^{m}$. Let  $P(z)=a_{0}+a_{1}\,z+\dots
+a_{n}\,z^{n}$ be a polynomial. We will denote by $\bar{P}$ the
polynomial $\bar{P}(z)=a_{n}+a_{n-1}\,z+\dots+ a_{0}\,z^{n}$ and we
define $P^{\frac{m}{n}}_{+}(z)=z^{m}
\left[\bar{P}^{\frac{m}{n}}\right]_{m}(z^{-1})$ for any positive
integer $m$. With these notations in place, we can now state the
following: 
\begin{prop}\label{prop-2}
The following two relations hold for any $n\geq 2$:
\begin{proplist}
\renewcommand{\theenumi}{\roman{enumi}}   
\item\label{prop-2-1} $\left[g_{n}^{\frac{n+1}{n}}\right]_{n-1}(z) =
\left[g_{n}^{\frac{n+1}{n}}\right]_{\lfloor\frac{n+1}{2}\rfloor}(z)
=g_{n+1}(z)$ 
\item\label{prop-2-2} $\left[g_{n}^{\frac{n-1}{n}}\right]_{n-2}(z) =
\left[g_{n}^{\frac{n-1}{n}}\right]_{\lfloor\frac{n-1}{2}\rfloor}(z)
=g_{n-1}(z)$ 
\end{proplist}
\end{prop}
\begin{proof}
For (\ref{prop-2-1}) notice that $g_{n}^{\frac{n+1}{n}}$ solves the
differential equation 
\begin{displaymath}
\frac{1}{n+1}\,g_{n}(z)\,\phi'(z)-\frac{1}{n}\,g_{n}'(z)\,\phi(z)=0
\end{displaymath}
with initial condition $\phi(0)=1$. From (\ref{prop-1-6}) of
Proposition \ref{prop-1} we see that $g_{n+1}$ solves the differential
equation 
\begin{equation}\label{eq-13}
\frac{1}{n+1}\,g_{n}(z)\,\phi'(z)-\frac{1}{n}\,g_{n}'(z)\,\phi(z)=(-z)^{n-1}
\end{equation}
with initial condition $\phi(0)=1$. It follows from this that
$\psi=g_{n+1}-g_{n}^{\frac{n+1}{n}}$ solves (\ref{eq-13}) with initial
condition $\phi(0)=\psi(0)=0$. If we now evaluate both sides in
(\ref{eq-13}) at $z=0$, with the identification $\phi=\psi$, we  get that
$\psi'(0)=0$. Differentiating (\ref{eq-13}) and evaluating at $z=0$ 
(again with the substitution $\phi=\psi$) we get $\psi''(0)=0$. 
Repeating the process $n-2$ times we get 
\begin{displaymath}
\psi(0)=\psi^{(1)}(0)=\dots=\psi^{(n-1)}(0)=0
\end{displaymath}
It follows from this that
\begin{displaymath} 
g_{n+1}(z)=\left[g^{\frac{n+1}{n}}\right]_{n-1}(z)
\end{displaymath}
Since $g_{n+1}$ has degree $\lfloor\frac{n+1}{2}\rfloor$ and
$\lfloor\frac{n+1}{2}\rfloor\leq n-1$ when $n\geq 2$, it follows that  
\begin{displaymath}
\left[g^{\frac{n+1}{n}}\right]_{n-1}(z) =
\left[g^{\frac{n+1}{n}}\right]_{\lfloor\frac{n+1}{2}\rfloor}(z)
\end{displaymath}
The proof for (\ref{prop-2-2}) is similar.
\end{proof}
\begin{prop}\label{prop-3}
For any integers $n\geq 3$ and $k\geq 0$ let $r_{n,k}$ be the
coefficient of $z^{kn+2}$ in the Taylor expansion of
$(g_{n}(z^2))^{k+\frac{1}{n}}$ around $z=0$ and $\tilde{r}_{n,k}$ the
coefficient of $z^{kn}$ in the expansion of
$(g_{n}(z^2))^{k-\frac{1}{n}}$. Then, 
\begin{displaymath}
r_{n,k}=\left\{\begin{array}{ll}\frac{(-1)^{nl}}{n^{l}l!}\,
\prod\limits_{m=l+1}^{2\,l}\,(m\,n+1),&k=2\,l\\ & \\0,&k=2\,l+1\end{array}\right.
\end{displaymath}
and
\begin{displaymath}
\tilde{r}_{n,k}=\left\{\begin{array}{ll}\frac{(-1)^{nl}}{n^{l}l!}\,
\prod\limits_{m=l+1}^{2\,l}\,(m\,n-1),&k=2\,l\\ & \\0,&k=2\,l+1\end{array}\right.
\end{displaymath}
\end{prop}
\begin{proof}
We are only going to prove the formula for $r_{n,k}\,$, since the
proof for $\tilde{r}_{n,k}$ is almost identical. 

For $k=0$, it is easy to see (using (\ref{prop-1-4}) of Proposition 1) that
\begin{displaymath}
r_{n,0}=\frac{d}{dz}\left(g_{n}^{\frac{1}{n}}(z)\right)_{z=0}=
\frac{1}{n}\,n\,h_{n-2}(0)\,\left(g_{n}^{\frac{1}{n}-1}(0)\right)=1
\end{displaymath}

For $k=1$, $r_{n,1}$ is the coefficient of $z^{n+2}$ in the Taylor
expansion of $g_{n}^{1+\frac{1}{n}}(z^2)$. If $n$ is odd this is
obviously $0$, because the Taylor series contains only even powers. If
$n$  is even, namely $n=2m\,$, then $r_{2m,1}$ is equal to the
coefficient of $z^{m+1}$ in the Taylor series of
$g_{2m}^{1+\frac{1}{2m}}(z)$, which is $0$ according to Proposition
\ref{prop-2}, if $n\geq 3$. Hence, for any $n\geq 3\,$, $r_{n,1}=0$.  

From the residue theorem and the fact that
$g_{n}(z^2)=z^n\,l_{n}(z^{-1})$ it follows that: 
\begin{eqnarray}\label{eq-14}
&r_{n,k}&=\frac{1}{2\pi i}\int_{\Gamma}\,\frac{1}{z^{kn+3}}\,
\left(z^{n}\,l_{n}(z^{-1})\right)^{k+\frac{1}{n}}\,dz\nonumber\\
&&=\frac{1}{2\pi
i}\int_{\Gamma}\,\frac{1}{z^{(k-1)n+3}}\,l_{n}(z^{-1})\,
\left(z^{n}\,l_{n}(z^{-1})\right)^{k-1+\frac{1}{n}}\,dz\nonumber\\
&&=\frac{1}{2\pi
i}\int_{\Gamma}\,\frac{1}{z^{(k-2)n+3}}\,l^{2}_{n}(z^{-1})\,
\left(z^{n}\,l_{n}(z^{-1})\right)^{k-2+\frac{1}{n}}\,dz
\end{eqnarray}
where the contour of integration $\Gamma$ is contained in a
neighborhood of $z=0$ where ($g_{n}(z^2))^{k+\frac{1}{n}}$
is analytic. From (\ref{eq-14}) we obtain after integrating by parts: 
\begin{eqnarray*}
&r_{n,k}&=-\frac{1}{kn+2}\,\frac{1}{2\pi i}\int_{\Gamma}\,\frac{d}{dz}
\left(\frac{1}{z^{kn+2}}\right)\,\left(z^{n}\,l_{n}(z^{-1})\right)^
{k+\frac{1}{n}}\,dz\\
&&=\frac{kn+1}{kn+2}\,\frac{1}{2\pi
i}\int_{\Gamma}\,\frac{1}{z^{(k-1)n+3}}\,
\left(l_{n}(z^{-1})+z^{-1}\,f_{n}(z^{-1})\right)\,\left(z^{n}\,l_{n}(z^{-1})
\right)^{k-1+\frac{1}{n}}\,dz\\
&&=\frac{2kn+2}{kn+2}\,\frac{1}{2\pi
i}\int_{\Gamma}\,\frac{1}{z^{(k-1)n+3}}\,
f_{n-1}(z^{-1})\,\left(z^{n}\,l_{n}(z^{-1})\right)^{k-1+\frac{1}{n}}\,dz
\end{eqnarray*}
where we have also used (\ref{eq-5}) and (\ref{eq-3}). We can rewrite
this last result as: 
\begin{equation}\label{eq-15}
\frac{kn+2}{2kn+2}\,r_{n,k}=\frac{1}{2\pi i}\int_{\Gamma}\,
\frac{1}{z^{(k-1)n+3}}\,f_{n-1}(z^{-1})\,\left(z^{n}\,l_{n}(z^{-1})\right)^
{k-1+\frac{1}{n}}\,dz
\end{equation}
Moreover, it also follows from (\ref{eq-14}) that
\begin{eqnarray*}
&r_{n,k}&=\frac{1}{2\pi i}\int_{\Gamma}\,\frac{1}{z^{(k-1)n+3}}\,
l_{n}(z^{-1})\,\left(z^{n}\,l_{n}(z^{-1})\right)^{k-1+\frac{1}{n}}\,dz\\
&&=\frac{1}{2\pi i}\int_{\Gamma}\,\frac{1}{z^{(k-1)n+3}}\,
\left(f_{n+1}(z^{-1})+f_{n-1}(z^{-1})\right)\,\left(z^{n}\,l_{n}(z^{-1})
\right)^{k-1+\frac{1}{n}}\,dz\\
&&=\frac{kn+2}{2kn+2}\,r_{n,k}+\frac{1}{2\pi i}\int_{\Gamma}\,
\frac{1}{z^{(k-1)n+3}}\,f_{n+1}(z^{-1})\,\left(z^{n}\,l_{n}(z^{-1})
\right)^{k-1+\frac{1}{n}}\,dz
\end{eqnarray*}
and, therefore, we have
\begin{equation}\label{eq-16}
\frac{kn}{2kn+2}\,r_{n,k}=\frac{1}{2\pi i}\int_{\Gamma}\,
\frac{1}{z^{(k-1)n+3}}\,f_{n+1}(z^{-1})\,\left(z^{n}\,l_{n}(z^{-1})
\right)^{k-1+\frac{1}{n}}\,dz
\end{equation}
Now if we multiply (\ref{eq-15}) by $n-1$ and (\ref{eq-16}) by $n+1$
and add them we get: 
\begin{eqnarray*}
&&\frac{kn^2+n-1}{kn+1}\,r_{n,k}=\frac{1}{2\pi i}\int_{\Gamma}\,
\frac{1}{z^{(k-1)n+3}}\,\left((n+1)\,f_{n+1}(z^{-1})+(n-1)\,f_{n-1}(z^{-1})
\right)\times\\
&&\mbox{}\times\left(z^{n}\,l_{n}(z^{-1})\right)^{k-1+\frac{1}{n}}\,dz\\
&&=\frac{1}{2\pi i}\int_{\Gamma}\,\frac{1}{z^{(k-1)n+1}}\,\frac{d}{dz}
\left(l_{n+1}(z^{-1})+l_{n-1}(z^{-1})\right)\,\left(z^{n}\,l_{n}(z^{-1})
\right)^{k-1+\frac{1}{n}}\,dz\\
&&=-\frac{1}{2\pi
i}\int_{\Gamma}\,\left(l_{n+1}(z^{-1})+l_{n-1}(z^{-1})
\right)\,\frac{d}{dz}\left(\frac{1}{z^{(k-1)n+1}}\,\left(z^{n}\,
l_{n}(z^{-1})\right)^{k-1+\frac{1}{n}}\right)\,dz\\
&&=-((k-1)n+1)\,\frac{1}{2\pi
i}\int_{\Gamma}\,\frac{1}{z^{(k-1)n+2}}\,
\left(z^{-2}+4\right)\,f_{n}(z^{-1})\,\left(z^{n}\,l_{n}(z^{-1})
\right)^{k-1+\frac{1}{n}}\,dz\\
&&\mbox{}+((k-1)n+1)\,\frac{1}{2\pi
i}\int_{\Gamma}\,\frac{1}{z^{(k-2)n+2}}\,
\left(z^{-2}+4\right)\,f_{n}(z^{-1})\,l_{n}(z^{-1})\,\left(z^{n}\,
l_{n}(z^{-1})\right)^{k-2+\frac{1}{n}}\,dz\\
&&\mbox{}-((k-1)n+1)\,\frac{1}{2\pi
i}\int_{\Gamma}\,\frac{1}{z^{(k-2)n+3}}\,
\left(z^{-2}+4\right)\,f^{2}_{n}(z^{-1})\,\left(z^{n}\,l_{n}(z^{-1})
\right)^{k-2+\frac{1}{n}}\,dz
\end{eqnarray*}
so that
\begin{eqnarray}\label{eq-17}
\frac{kn^2+n-1}{(kn+1)((k-1)n+1)}\,r_{n,k}&=&-\frac{1}{2\pi
i}\int_{\Gamma}\,
\frac{1}{z^{(k-2)n+3}}\,\left(z^{-2}+4\right)\,f^{2}_{n}(z^{-1})\nonumber\\
&=&\mbox{}\times\left(z^{n}\,l_{n}(z^{-1})\right)^{k-2+\frac{1}{n}}\,dz
\end{eqnarray}
Finally, adding (\ref{eq-16}) and (\ref{eq-17}) we get:
\begin{eqnarray*}
4\,(-1)^{n}\,r_{n,k-2}&=&\frac{1}{2\pi i}\int_{\Gamma}\,
\frac{1}{z^{(k-2)n+3}}\,\left[l_{n}(z^{-1})-\left(z^{-2}+4\right)\,
f_{n}^{2}(z^{-1})\right]\\
 &  & \times\left(z^{n}\,l_{n}(z^{-1})\right)^
{k-2+\frac{1}{n}}\,dz\\
&=&\frac{kn\,(kn+2)}{(kn+1)\,((k-1)n+1)}\,r_{n,k}
\end{eqnarray*}
In other words:
\begin{equation}\label{eq-18}
r_{n,k}=4\,(-1)^{n}\,\frac{(kn+1)\,((k-1)n+1)}{kn\,(kn+2)}\,r_{n,k-2}
\end{equation}
and the desired result follows by induction.
\end{proof}

\section{The Lax function and the Lax equation:}
In this section we show that, for $n\geq 3$, the Lax function 
\begin{equation}\label{eq-19}
L_{n}=\frac{(\sqrt{-v})^{n}}{n}\,l_{n}\left(\frac{n^{\frac{1}{n}}\,p}
{\sqrt{-v}}\right)+u
\end{equation}
provides a standard Lax description for the polytropic gas with
parameter  $\gamma=n+1$, namely
\begin{equation}\label{eq-20}
\frac{\partial\,L_{n}}{\partial\,t}=\frac{n}{n+1}\,\left\{\left(
L_{n}^{1+\frac{1}{n}}\right)_{+}\,,\,L_{n}\right\}
\end{equation}
where $()_{+}$ denotes the part of the polynomial that contains only
non-negative powers of the variable, leads to the  dynamical equations
of  motion
\begin{eqnarray}
&&u_{t}+u\,u_{x}+v^{\gamma-2}\,v_{x}=0\label{eq-21}\\
&&v_{t}+(u\,v)_{x}=0\label{eq-22}
\end{eqnarray}
which are the polytropic gas equations. 

First, let us introduce the following simplifying notations:
\begin{displaymath}
\begin{array}{l}
w=-v,\quad\pi_{n}=n^{\frac{1}{n}}\,p,\quad\tilde{L}_{n}=L_{n}-u\\
\left\{A,B\right\}_{n}=\frac{\partial\,A}{\partial\,w}\,\frac{\partial\,B}
{\partial\,\pi_{n}}-\frac{\partial\,A}{\partial\,\pi_{n}}\,
\frac{\partial\,B}{\partial\,w}=\left(n^{\frac{1}{n}}\,w_{x}\right)^{-1}\,
\left\{A,B\right\}
\end{array}
\end{displaymath}
where $\{A,B\}$ denotes the conventional Poisson bracket of
$A,B$. With these  notations, we can now rewrite (\ref{eq-19}) as
\begin{equation}\label{eq-23}
L_{n} = \frac{1}{n}\,\left(\sqrt{w}\right)^{n}\,l_{n}\left(\left(\sqrt{w}
\right)^{-1}\,\pi_{n}\right)+u=\frac{1}{n}\,\pi_{n}^{n}\,g_{n}
\left(w\,\pi_{n}^{-2}\right)+u
\end{equation}
From this we see that 
\begin{eqnarray}\label{eq-24}
\left(L_{n}^{1+\frac{1}{n}}\right)_{+}&=&\left(\frac{1}{n}\right)^
{1+\frac{1}{n}}\,\pi_{n}^{n+1}\,\left[\left(g_{n}(w\,z^2)+n\,u\,z^{n}\right)^
{1+\frac{1}{n}}\right]_{n+1}(z=\pi_{n}^{-1})\nonumber\\
&=&\left(\frac{1}{n}\right)^{1+\frac{1}{n}}\,\pi_{n}^{n+1}\,\left(\left[
\left(g_{n}(w\,z^{2})\right)^{1+\frac{1}{n}}\right]_{n+1}(z=\pi_{n}^{-1})
+(n+1)\,u\,\pi_{n}^{-n}\right)\nonumber\\
&=&\left(\frac{1}{n}\right)^{1+\frac{1}{n}}\,\pi_{n}^{n+1}\,\left(\left[
\left(g_{n}(z)\right)^{1+\frac{1}{n}}\right]_{\lfloor\frac{n+1}{2}\rfloor}
(z=w\,\pi_{n}^{-2}) +(n+1)\,u\,\pi_{n}^{-n}\right)\nonumber\\
&=&\left(\frac{1}{n}\right)^{1+\frac{1}{n}}\,\pi_{n}^{n+1}\,g_{n+1}
\left(w\,\pi_{n}^{-2}\right)+(n+1)\,\left(\frac{1}{n}\right)^
{1+\frac{1}{n}}\,u\,\pi_{n}\nonumber\\
&=&\left(\tilde{L}_{n}^{1+\frac{1}{n}}\right)_{+}+(n+1)\,\left(\frac{1}{n}
\right)^{1+\frac{1}{n}}\,u\,\pi_{n}\nonumber\\
 &=&\left(\tilde{L}_{n}^{1+\frac{1}{n}}
\right)_{+}+\frac{n+1}{n}\,u\,p
\end{eqnarray}
where we have used Proposition \ref{prop-2}. Using the above relation,
the Poisson bracket on the right hand side of the Lax equation
(\ref{eq-20}) becomes: 
\begin{eqnarray}\label{eq-25}
\left\{\left(L_{n}^{1+\frac{1}{n}}\right)_{+},\;L_{n}\right\}&=&\left\{
\left(\tilde{L}_{n}^{1+\frac{1}{n}}\right)_{+},\;\tilde{L}_{n}\right\}-
u_{x}\,\frac{\partial}{\partial\,p}\,\left(\tilde{L}_{n}^{1+\frac{1}{n}}
\right)_{+}\nonumber\\
&&\mbox{}+\frac{n+1}{n}\left(u_{x}\,p\,\frac{\partial\,\tilde{L}_{n}}
{\partial\,p}-u\,\frac{\partial\,\tilde{L}_{n}}{\partial\,x}-u\,u_{x}\right)
\end{eqnarray}
Now, with (\ref{prop-1-5}) of Proposition \ref{prop-1} we obtain:
\begin{eqnarray}\label{eq-26}
\frac{\partial}{\partial\,p}\,\left(\tilde{L}_{n}^{1+\frac{1}{n}}\right)_{+}
&=&\frac{\partial\,\pi_{n}}{\partial\,p}\,\frac{\partial}
{\partial\,\pi_{n}}\,\left(n^{-\left(1+\frac{1}{n}\right)}\,\pi_{n}^{n+1}\,
g_{n+1}(w\,\pi_{n}^{-2})\right)\nonumber\\
&=&\frac{n+1}{n}\,\pi_{n}^{n}\,\left(g_{n+1}(w\,\pi_{n}^{-2})-\frac{2}{n+1}\,
(w\,\pi_{n}^{-2})\,g_{n+1}'(w\,\pi_{n}^{-2})\right)\nonumber\\
&=&\frac{n+1}{2\,n}\,\left(2\,\pi_{n}^{n}\,g_{n}(w\,\pi_{n}^{-2})-
\frac{2}{n}\,w\,\pi_{n}^{n-2}\,g_{n}'(w\,\pi_{n}^{-2})\right)\nonumber\\
&=&\frac{n+1}{2\,n}\,\left(\pi_{n}\,\frac{\partial\,\tilde{L}_{n}}
{\partial\,\pi_{n}}+n\,\tilde{L}_{n}\right)\nonumber\\
&=&\frac{n+1}{2\,n}\,\left(p\,\frac{\partial\,\tilde{L}_{n}}
{\partial\,p}+n\,\tilde{L}_{n}\right)\nonumber\\
\end{eqnarray}
Furthermore, using (\ref{prop-1-6}) of the same proposition we
notice that: 
\begin{eqnarray}\label{eq-27}
\left\{\left(\tilde{L}_{n}^{1+\frac{1}{n}}\right)_{+},\;\tilde{L}_{n}\right\}
&=&\left(n^{\frac{1}{n}}\,w_{x}\right)\,\left\{\left(\tilde{L}_{n}^{1+
\frac{1}{n}}\right)_{+},\;\tilde{L}_{n}\right\}_{n}\nonumber\\
&=&n^{-2}\,w_{x}\,\left\{\pi_{n}^{n+1}\,g_{n+1}(w\,\pi_{n}^{-2}),\;
\pi_{n}^{n}\,g_{n}(w\,\pi_{n}^{-2})\right\}_{n}\nonumber\\
&=&\frac{n+1}{n}\,w_{x}\,\pi_{n}^{2\,n-2}\,\left(\frac{1}{n+1}\,g_{n+1}'
(w\,\pi_{n}^{-2})\,g_{n}(w\,\pi_{n}^{-2})-\right.\nonumber\\
&&\left.\mbox{}-\frac{1}{n}\,g_{n+1}(w\,\pi_{n}^{-2})\,g_{n}'
(w\,\pi_{n}^{-2})\right)\nonumber\\
&=&-\frac{n+1}{n}\,v^{n-1}\,v_{x}
\end{eqnarray}
Putting together (\ref{eq-25}), (\ref{eq-26}) and (\ref{eq-27}) we see
that the Lax equation (\ref{eq-20}) is equivalent to 
\begin{equation}\label{eq-28}
\frac{\partial\,L_{n}}{\partial\,t}=\frac{1}{2}\left(p\,\frac{\partial\,L_{n}}
{\partial\,p}-n\,\left(L_{n}-u\right)\right)\,u_{x}-\frac{\partial\,L_{n}}
{\partial\,x}\,u-v^{n-1}\,v_{x}
\end{equation}
Equating coefficients of powers of $p$ in (\ref{eq-28}) we see that it
holds if and only if the equations of motion (\ref{eq-21}) and
(\ref{eq-22}) hold.

This shows that the Lax function in (\ref{eq-19}) does provide a
standard Lax description for the polytropic gas dynamics. The same Lax
function also provides a standard Lax description for the elastic
medium equations, which we show in the appendix. For $\gamma = 3\,
(n=2)$, the polytropic gas equations, under a redefinition of
variables, is known to describe two decoupled Riemann equations. This
case is slightly tricky, since in this case, both the dynamical
variables, $u,v$ have the same dimension. We describe the Lax
description for  this separately, which does not fall into the above
category. Consider the Lax function

\begin{equation}
L = p^{2} + {2\over 3} u + {1\over 6} v^{2} p^{-2}
\end{equation}
The presence of the $p^{-2}$ gives it a different character from the
earlier construction. However, it is straightforward to check that the
Lax equation

\begin{equation}
{\partial L\over \partial t} = \left\{\left(L^{3\over 2}\right)_{+} ,
L\right\} 
\end{equation}
leads to

\begin{eqnarray}
u_{t} & = & - u u_{x} - v v_{x}\nonumber\\
v_{t} & = & - (uv)_{x}
\end{eqnarray}
which are the polytropic gas equations for $\gamma =3$. The two sets
of conserved charges, in this case, are obtained from the two possible
ways of defining the residues of $L^{n+{1\over 2}}$ around
$p=0,\infty$ respectively. We also note here that, for the case
$\gamma =2\,(n=1)$, the polytropic gas equations only have a
nonstandard Lax description.

\section{Conserved charges:}
A Lax description of an integrable system has the advantage that the
conserved charges can be obtained from residues of fractional powers
of the Lax function. The polytropic gas, on the other hand, is known
to have two infinite sets of conserved charges \cite{9}. In the
non-standard Lax description of the polytropic gas, it is known that
the two sets of charges arise naturally from calculating the residues
around two distinct points. In a standard Lax description, however,
the residues are unique. It is
interesting, therefore, to see how the two sets of conserved charges
will arise in this description. Let us note that a simple dimensional
analysis  shows that we can assign the dimensions $[v]=2, [u] = n$ so
that $[L_{n}] = n$. From the explicit forms of the two sets of
known conserved charges \cite{9}, it is clear that  if they
are obtained from our Lax function at all, they should arise from the
fractional
powers  $k+\frac{1}{n}$ and $k-\frac{1}{n}$ respectively, where
$k=0,1,\cdots$ for the first set while $k=1,2,\cdots$ for the second set. 

Let $H_{n,k}=\mathrm{Res}\left(L_{n}^{k+\frac{1}{n}}\right)$ and
$F_{n,k}(u,w,z)=\left(g_{n}(w\,z^{2})+n\,u\,z^{n}\right)^{k+\frac{1}{n}}$.
Let $\rho_{n,k}(u,w)$ be the coefficient of $z^{kn+2}$ in the Taylor
expansion of $F_{n,k}$ around $z=0$. Then: 
\begin{equation}\label{eq-29}
H_{n,k}=n^{-(k+\frac{2}{n})}\rho_{n,k}(u,w)
\end{equation}
Since 
\begin{equation}\label{eq-30}
\frac{\partial\,F_{n,k}}{\partial\,u}(u,w,z)=(k\,n+1)\,z^{n}\,F_{n,k-1}(u,w,z)
\end{equation}
it follows that 
\begin{equation}\label{eq-31}
\frac{\partial\,\rho_{n,k}}{\partial\,u}(u,w)=\left\{\begin{array}{ll}0&
\mathrm{if}\;k=0\\(k\,n+1)\rho_{n,k-1}(u,w)&\mathrm{if}\;k>0\end{array}\right.
\end{equation}
and, after repeated integration:
\begin{equation}\label{eq-32}
\rho_{n,k}(u,w)=\sum_{m=0}^{k}\,\frac{1}{m!}\,
\left(\prod_{l=k-m+1}^{k}\,l\,n+1\right)\,\rho_{n,k-m}(0,w)\,u^{m}
\end{equation}
Using the fact that $\rho_{n,2m}(0,w)=r_{n,2m}\,w^{m\,n+1}$ in
conjunction with Proposition \ref{prop-3} we get: 
\begin{equation}\label{eq-33}
\rho_{n,k}(u,w)=\sum_{m=0}^{\lfloor\frac{k}{2}\rfloor}\,\frac{(-1)^{m\,n}}
{m!\,(k-2\,m)!\,n^{m}}\,\left(\prod_{l=m+1}^{k}\,l\,n+1\right)\,w^{mn+1}\,
u^{k-2m}
\end{equation}
and therefore
\begin{equation}\label{eq-34}
H_{n,k}=-n^{-(k+\frac{2}{n})}\,\prod_{s=0}^{k}\,(s\,n+1\,)
\sum_{m=0}^{\lfloor\frac{k}{2}\rfloor}\frac{1}{m!\,(k-2\,m)!\,n^{m}}\,
\prod_{l=0}^{m}\,{1\over l\,n+1}\,v^{mn+1}u^{k-2m}
\end{equation}
Similarly, if
$\tilde{H}_{n,k}=\mathrm{Res}\,\left(L_{n}^{k-\frac{1}{n}}\right)$
then we can show that 
\begin{equation}
\tilde{H}_{n,k}=n^{-k}\,\prod_{s=0}^{k}(s\,n-1\,)
\sum_{m=0}^{\lfloor\frac{k}{2}\rfloor}\frac{1}{m!\,(k-2\,m)!\,n^{m}}
\prod_{l=0}^{m}{1\over l\,n -1}\,v^{mn}\,u^{k-2m}
\end{equation}
The two sets of conserved charges simply correspond to $\int
H_{n,k}\,dx$ and $\int\tilde{H}_{n,k}\,dx$ and coincide with the
known conserved charges constructed earlier from a nonstandard Lax
representation upto an overall normalization. This construction shows
that  the two sets of conserved
charges can be obtained from the residues of two distinct families of
fractional powers of the Lax function in this standard description.

\section{Generalization to dispersive cases:}

It is an interesting question to ask which dispersive integrable models
reduce in the dispersionless limit to the polytropic gas dynamics. It
is, of course, well known that the two boson equation, in the
dispersionless limit, goes to the polytropic gas equation for $n = 1\,
(\gamma =2)$  and
that, for $n =2\, (\gamma =3)$, the polytropic gas equation is
equivalent to two
decoupled Riemann equations which can be thought of as the
dispersionless limit of the KdV equation. It is, of course, clear that
there may be several dispersive models whose dispersionless limit will
give the same equation. However, our interest is to find even one
family of such models. Surprisingly, beyond $n=4$, we have not found
any dispersive generalization of these systems.

It is well known that the polytropic gas equations for $\gamma =2\,
(n=1)$ can be thought of as the dispersionless limit of the two boson
equation, which is described by the Lax operator

\begin{equation}
L = \partial - u + \partial^{-1} v
\end{equation}
and the nonstandard Lax equation

\begin{equation}
{\partial L\over \partial t} = {1\over
2}\left[\left(L^{2}\right)_{\geq 1} , L \right]
\end{equation}
We will not discuss this system any further.

The case $n=2$ is tricky even at the dispersionless limit, as we have
already pointed out. We note that the Riemann equation is well
understood to be the dispersionless limit of the KdV equation, whose
Lax description is one of the first to have been studied
\cite{14}. However, for
$n=2$, we have two decoupled Riemann equations in the dispersionless
limit. The dispersive generalization of this system is not at all
clear. However, with some work, we have found that the Lax operator

\begin{equation}
L = \partial^{2} + {1\over 2} u + {1\over 2} \partial^{-1} u \partial
+ {1\over 4} \partial^{-1} v \partial^{-1} v
\end{equation}
through the nonstandard Lax equation

\begin{equation}
{\partial L\over \partial t} = -{2\over 3} \left[\left(L^{3\over
2}\right)_{\geq 1} , L \right]
\end{equation}
leads to

\begin{eqnarray}
u_{t} & = & - {2\over 3} u_{xxx} - u u_{x} - v v_{x}\nonumber\\
v_{t} & = & - {2\over 3} v_{xxx} - (uv)_{x}
\end{eqnarray}
With a simple redefinition of variables to $ u\pm v$, this becomes two
decoupled KdV equations, which, therefore, provides a trivial
dispersive  generalization of the
polytropic gas equations for $\gamma = 3$ (or $n=2$). It is worth
pointing out here that this Lax operator also gives through

\begin{equation}
{\partial L\over \partial t} = - \left[\left(L^{1\over 2}\right)_{\geq
1} , L\right]
\end{equation}
leads to

\begin{equation}
u_{t} = - u_{x},\qquad v_{t} = - v_{x}
\end{equation}
which are the elastic medium equations for $\gamma =3$. Under this
description, they do not seem to pick up any dispersive terms.

For $\gamma = 4\, (n=3)$, we expect the dispersive generalization to be
related to the Boussinesq hierarchy, simply from the counting of
dimensions. In fact, if we choose

\begin{equation}
L = \partial^{3} + v \partial + u
\end{equation}
the standard Lax equation

\begin{equation}
{\partial L\over \partial t} = \left[\left(L^{4\over 3}\right)_{+} , L
\right]
\end{equation}
leads to

\begin{eqnarray}
u_{t} & = & {1\over 9}\left(3u_{xxxx} - 2v_{xxxxx} - 6v_{xxx}v - 12
v_{xx}v_{x} + 6(vu_{x})_{x} + 12 uu_{x} - -
4v^{2}v_{x}\right)\nonumber\\
v_{t} & = & {1\over 3} \left(2u_{xxx} - v_{xxxx} - (v^{2})_{xx} +
4(uv)_{x}\right)
\end{eqnarray}
These clearly provide a dispersive generalization of the polytropic
gas equations for $\gamma =4$. These equations are known to admit the
$W_{3}$ algebra as one of the Hamiltonian structures.

Let us further note that the Lax equation

\begin{equation}
{\partial L\over \partial t} = \left[\left(L^{2\over 3}\right)_{+} ,
L\right]
\end{equation}
leads to the set of equations

\begin{eqnarray}
u_{t} & = & {1\over 3} \left(3u_{xx} - 2v_{xxx} -
2vv_{x}\right)\nonumber\\
v_{t} & = & -v_{xx} + 2u_{x}
\end{eqnarray}
which provides a dispersive generalization of the elastic medium
equations for $\gamma =4$. In general, we note that the equations
\begin{equation}
{\partial L\over \partial t} = \left[\left(L^{k\over 3}\right)_{+} ,
L\right]
\end{equation}
with $k = 3m \pm 1$ defines consistent equations corresponding to the
two different hierarchies.

For $\gamma =5\, (n=4)$, consider the Lax operator \cite{15}

\begin{equation}
L = \partial^{4} + v \partial^{2} + v_{x}\partial + u
\end{equation}
This leads to consistent equations through

\begin{equation}
{\partial L\over \partial t} = \left[\left(L^{k\over 4}\right)_{+} ,
L\right]
\end{equation}
for $k = 4m \pm 1$. Thus, for $k = 5$, the dynamical equations turn
out to be
\begin{eqnarray}
u_{t} & = & {1\over 32}\left( 12 u_{xxxxx} - 5v_{xxxxxxx} -
5(2vv_{xxxx} + 4v_{x}v_{xxx} + 3v_{xx}^{2})_{x} + 20 (u_{x}v)_{xx}
\right.\nonumber\\
 &  & \quad \left. + 40 uu_{x} - 5 (v^{2}v_{xx} + vv_{x}^{2})_{x} +
5v^{2}u_{x}\right)\nonumber\\
v_{t} & = & {1\over 32} \left(40u_{xxx} - 18 v_{xxxxx} - 15 (2vv_{xx}
+ v_{x}^{2})_{x} + 40 (uv)_{x} - 15 v^{2}v_{x}\right)
\end{eqnarray}
With a simple change of variables, $\tilde{u} = u - {1\over 8} v^{2},
\tilde{v} = v$, it is easy to check that these equations reduce, in the
dispersionless limit, to the polytropic gas equations with $\gamma =
5$. Therefore, this model provides a dispersive generalization of these
equations. It is worth noting here that this system of
equations admits as a Hamiltonian structure (after suitable redefinition
of fields) the nonlinear $W$ algebra, $W(2,4)$, which is uniquely
characterized by the presence of a spin 2 Virasoro field as well as a
spin 4 primary field.

Similarly, it can be easily checked that, for $k=3$, the Lax equation
leads to
\begin{eqnarray}
u_{t} & = & u_{xxx} - {3\over 8} v_{xxxxx} - {3\over 8} (vv_{xx})_{x}
+ {3\over 4} u_{x}v\nonumber\\
v_{t} & = & - {5\over 4} v_{xxx} + 3u_{x} - {3\over 4} vv_{x}
\end{eqnarray}
which, under the same redefinition of variables, goes over to the elastic
medium equations for $\gamma =5$ in the dispersionless limit. This
system of equations, therefore, gives a dispersive generalization of
both these systems. 

For $\gamma =6\, (n=5)$ as well as $\gamma =7\,(n=6)$, we have
explicitly  verified that there is no
standard Lax equation that leads to consistent equations with two
dynamical variables. For higher values of $n$ (and, therefore, $\gamma$),
it is unlikely that a standard Lax description would lead to
consistent equations with two dynamical fields, since the number of
consistency conditions increases rapidly. However, we have not checked
this explicitly beyond $n=6$. The dispersive generalization of
the polytropic gas equations for higher values of $\gamma$, therefore,
remains an open question. It is possible that they arise only as
nonstandard equations or that one may have to introduce additional dynamical
variables, which, somehow, disappear in the dispersionless limit.

\section{Conclusion:}

We have derived a standard Lax description for the polytropic gas
dynamics. The Lax function, in this case, is intimately connected with
Lucas polynomials, which are also related to the Fibonacci
polynomials. The two infinite sets of conserved charges have been
obtained from the residues of two distinct sequences of fractional
powers of the Lax function. We have  shown that the same Lax
function also provides a standard Lax description for the elastic
medium equations. In addition, we have presented some results on
possible dispersive generalizations of such systems. 

This work was supported in part by US Deparment of Energy grant number 
DE-FG-02-91ER40685 and by CNPq-Brasil.
\appendix
\section{Standard Lax description for elastic medium equations:}
In this appendix, we show how the same Lax function of (\ref{eq-19})
leads to a standard Lax description for the elastic medium equations
\cite{7,9}. 
\begin{eqnarray*}
\left(L_{n}^{1-\frac{1}{n}}\right)_{+}&=&n^{\frac{1}{n}-1}\,\pi_{n}^{n-1}\,\left[\left(g_{n}(w\,z^{2})+n\,u\,z^{n}\right)^{1-\frac{1}{n}}\right]_{n-1}(z=\pi_{n}^{-1})\\
&=&n^{\frac{1}{n}-1}\,\pi_{n}^{n-1}\,\left[\left(g_{n}(w\,z^{2})\right)^{1-\frac{1}{n}}\right]_{n-1}(z=\pi_{n}^{-1})\\
&=&n^{\frac{1}{n}-1}\,\pi_{n}^{n-1}\,\left[\left(g_{n}(z)\right)^{1-\frac{1}{n}}\right]_{\lfloor\frac{n-1}{2}\rfloor}(z=w\,\pi_{n}^{-2})\\
&=&n^{\frac{1}{n}-1}\,\pi_{n}^{n-1}\,g_{n-1}(w\,\pi_{n}^{-2})=\left(\tilde{L}_{n}^{1-\frac{1}{n}}\right)_{+}
\end{eqnarray*}
\begin{eqnarray*}
\left\{\left(L_{n}^{1-\frac{1}{n}}\right)_{+},\ L_{n}\right\}&=&\left\{\left(\tilde{L}_{n}^{1-\frac{1}{n}}\right)_{+},\ \tilde{L}_{n}\right\}-u_{x}\,\frac{\partial}{\partial\,p}\,\left(\tilde{L}_{n}^{1-\frac{1}{n}}\right)_{+}
\end{eqnarray*}
\begin{eqnarray*}
\frac{\partial}{\partial\,p}\,\left(\tilde{L}_{n}^{1-\frac{1}{n}}\right)_{+}&=&\frac{\partial\,\pi_{n}}{\partial\,p}\,\frac{\partial}{\partial\,\pi_{n}}\,\left(n^{\frac{1}{n}-1}\,\pi_{n}^{n-1}\,g_{n-1}(w\,\pi_{n}^{-2})\right)\\
&=&n^{\frac{2}{n}-1}\,(n-1)\,\pi_{n}^{n-2}\,\left(g_{n-1}(w\,\pi_{n}^{-2})-\frac{2}{n-1}\,(w\,\pi_{n}^{-2})g_{n-1}'(w\,\pi_{n}^{-2})\right)\\
&=&n^{\frac{2}{n}-1}\,(n-1)\,\pi_{n}^{n-2}\,\left(g_{n-2}(w\,\pi_{n}^{-2})-\frac{1}{n-2}\,(w\,\pi_{n}^{-2})g_{n-2}'(w\,\pi_{n}^{-2})\right)\\
&=&n^{\frac{2}{n}-1}\,(n-1)\,\left(\sqrt{w}\right)^{n-2}\,\left(l_{n-2}\left(\frac{\pi_{n}}{\sqrt{w}}\right)-f_{n-3}\left(\frac{\pi_{n}}{\sqrt{w}}\right)\right)\\
&=&n^{\frac{2}{n}-1}\,(n-1)\,\left(\sqrt{w}\right)^{n-2}\,f_{n-1}\left(\frac{\pi_{n}}{\sqrt{w}}\right)\\
&=&n^{\frac{2}{n}-1}\,(n-1)\,\pi_{n}^{n-2}\,h_{n-2}(w\,\pi_{n}^{-2})\\
&=&n^{\frac{2}{n}-1}\,(n-1)\,\frac{\partial\,\tilde{L}_{n}}{\partial\,w}
\end{eqnarray*}
\begin{eqnarray*}
\left\{\left(\tilde{L}_{n}^{1-\frac{1}{n}}\right)_{+},\ \tilde{L}_{n}\right\}&=&n^{\frac{1}{n}}\,w_{x}\,\left\{\left(\tilde{L}_{n}^{1-\frac{1}{n}}\right)_{+},\ \tilde{L}_{n}\right\}_{n}\\
&=&n^{2\,\left(\frac{1}{n}-1\right)}\,w_{x}\left\{\pi_{n}^{n-1}\,g_{n-1}(w\,\pi_{n}^{-2}),\ \pi_{n}^{n}(w\,\pi_{n}^{-2})\right\}_{n}\\
&=&n^{\frac{2}{n}-1}\,(n-1)\,w_{x}\,\pi_{n}^{2n-4}\nonumber\\
& & \times\left(\frac{1}{n-1}\,g_{n}(w\,\pi_{n}^{-2})\,g_{n-1}'(w\,\pi_{n}^{-2})-\frac{1}{n}\,g_{n}'(w\,\pi_{n}^{-2})\,g_{n-1}(w\,\pi_{n}^{-2})\right)\\
&=&(-1)^{n-1}\,n^{\frac{2}{n}-1}\,(n-1)\,w^{n-2}\,w_{x}
\end{eqnarray*}
Using these results, the Lax equation
\begin{displaymath}
\frac{\partial\,L_{n}}{\partial\,t}=-n^{1-\frac{2}{n}}\,\frac{1}{n-1}\,\left\{\left(L_{n}^{1-\frac{1}{n}}\right)_{+},\ L_{n}\right\}
\end{displaymath}
can be written
\begin{displaymath}
w_{t}\,\frac{\partial\,\tilde{L}_{n}}{\partial\,w}+u_{t}=u_{x}\frac{\partial\,\tilde{L}_{n}}{\partial\,w}+(-1)^{n}\,w^{n-2}\,w_{x}
\end{displaymath}
which holds if and only if
\begin{displaymath}
\left\{\begin{array}{l}v_{t}=-u_{x}\\u_{t}=-v^{n-2}\,v_{x}\end{array}\right.
\end{displaymath}
These are none other than the elastic medium equations. Thus, we
have shown that the same Lax function also provides a standard Lax
description for the elastic medium equations.

\end{document}